\documentstyle{article}
\title{Green's functions and propagation of waves in
strongly inhomogeneous media}
\author{ Z. Haba\\Institute of Theoretical Physics, University of Wroclaw,
\\50-204 Wroclaw, Plac Maxa Borna 9, Poland\\e-mail:zhab@ift.uni.wroc.pl}
\date{}
\begin{document}
\maketitle
\begin{abstract}
We show that Green's functions of second order differential
operators with singular or unbounded coefficients can have an
anomalous behaviour in comparison to the well-known properties of
the Green's functions of operators with bounded coefficients. We
discuss some consequences of such an anomalous short or long
distance behaviour for a diffusion and wave propagation in an
inhomogeneous medium.
\end{abstract}
\section{ Introduction}
A wave propagation with a constant speed is an approximation to
the real situation when the wave propagates in a medium with its
characteristics varying in space. A similar approximation is
applied when considering a diffusion. In general, the speed of the
diffusion can vary in space. If this variation is slow then one
could believe that its effects are negligible. It has been
observed some time ago ( see \cite{nature} and references quoted
there) that  a diffusion in  strongly inhomogeneous materials can
be anomalous. This happens in particular with a heat convection in
a turbulent medium \cite{frisch}\cite{gawedzki}. In general, it is
rather difficult to investigate these problems because exact
solutions are not available and approximations to equations with
varying coefficients are not reliable.We discuss here the proper
time method for a representation of the Green's functions. We
represent solutions of the equation
\begin{equation}
{\cal A}G_{E}=2\delta
\end{equation}
for the Green's function of an elliptic operator ${\cal A}$ in
terms of its heat kernel. In this way the Green's function is
expressed by a diffusion.
 A solution of such an equation allows to determine the wave
propagation if we consider equation (1)either as an analog of the
Helmholtz equation for the propagation of monochromatic waves
(then all the coordinates in (1) are spatial) or continue
analytically the solution $G_{E}$ in one coordinate (identified
with time) to imaginary values (we denote such a continuation by
$G_{F}$). Then, the Green's function $G_{F}$ is known as the
Feynman propagator. Its real part is a sum of the conventional
advanced and retarded propagators. In general, the Feynman
propagator is not relevant to classical field theory. However,
quantized electromagnetic field is necessary for a description of
photons moving in an active medium which can be described as a
dielectric or magnetic material
\cite{mandl}\cite{welsch}\cite{glauber} .

If the coefficients of ${\cal A}$ are bounded regular functions
then it is known \cite{hadamard}that there is a minor effect of
varying coefficients on the local behaviour of the Green's
functions. We show that if the coefficients can grow or are
singular then they have a profound effect on the  Green's
functions. Then, the change of the behaviour of the Green's
functions has impact on propagation of disturbances in the medium.

Let us mention some  applications of our results to the wave
equations in physics. Mechanical waves can be derived from Euler
and continuity equations \cite{waves}. For example, the wave
equation for the pressure $p$ reads
\begin{equation} \nu\partial_{t}^{2}p =\nabla\rho^{-1}\nabla p
\end{equation}
where $\nu$ is the compressibility and $\rho$ is the density . An
inhomogeneous equation whose solution is expressed by the Green's
function describes the effect  of external forces acting upon the
medium \cite{landau}. There is  an equation similar to eq.(2) for
the scalar potential $\phi$ defined by the velocity ${\bf
v}=-\nabla \phi$.

Next, we consider  electromagnetic waves. There is a remarkable
similarity between wave equations in electrodynamics and in fluid
dynamics. It follows from the Maxwell equations that \cite{yariv}
\begin{equation}
\epsilon\partial_{t}^{2}{\bf
E}=\nabla\mu^{-1}\times\nabla\times{\bf E}-\partial_{t}{\bf J}
\end{equation}
where $\mu$ is magnetic susceptibility and $\epsilon$ is the
dielectric permittivity. The equation for the magnetic field can
be written in the form
\begin{equation}
\mu\partial_{t}^{2}{\bf
H}=-\nabla\epsilon^{-1}\times\nabla\times{\bf
H}+\nabla\times\epsilon^{-1}{\bf J}
\end{equation}
We could also introduce  scalar $\phi$ and vector ${\bf A}$
potentials as \begin{displaymath} {\bf B}=\nabla\times {\bf A}
\end{displaymath}\begin{displaymath}
{\bf E}=-\nabla \phi-\partial_{t}{\bf A}\end{displaymath} Then, in
the static case it follows from Maxwell equations that the scalar
potential $\phi$ satisfies the equation
\begin{equation}
\partial_{j}\epsilon\partial_{j}\phi=-\rho
\end{equation}
where $\rho$ is the charge density. The vector potential ${\bf A}$
is a solution of the wave equation similar to eq.(4)
\cite{welsch}\cite{glauber}.

 The
operator on the rhs of eqs.(3)and (4) is mixing the components of
the electric  and magnetic fields. It takes a simpler form if
$\epsilon$, $\mu$ and ${\bf J}$  do not depend on one coordinate
$x_{D}$ and we are looking for solutions ($E_{D}$ or $H_{D}$)
which do not depend on $x_{D}$. Then, eq.(3) reads
\begin{equation}
\epsilon\partial_{t}^{2}E_{D}-\partial_{j}\mu^{-1}\partial_{j}E_{D}=-\partial_{t}J_{D}
\end{equation}
Taking the Fourier transform in time of eq.(6) we obtain the
Helmholtz equation
\begin{equation}
-\epsilon\omega^{2}E^{\omega}_{D}-\partial_{j}\mu^{-1}\partial_{j}E^{\omega}_{D}=-i\omega
J^{\omega}_{D}
\end{equation}
The quantum version of this model describes a quantum
electromagnetic field interacting with a classical source. The
correlation functions of the photon field and the scattering
matrix for photons scattered on the source (or produced by the
source) are described by the Feynman propagator. Eqs.(6)-(7) are
reduced to a two-dimensional space. In the next section we discuss
the scalar wave equation (2) in $D$ space-time dimensions ($D-1=d$
space dimension) for greater generality. The scalar wave equation
is often considered as a good approximation to the vector one. The
Poisson-type equation (5) can be considered as a special case of
eq.(6) when the time derivative is neglected. In this paper we
discuss quite unusual (unphysical) models for the space-dependent
coefficients in eqs.(2)-(7). However, our results show that the
conventional wisdom about spreading of forces in hydrodynamics and
electrodynamics may require a modification in strongly
inhomogeneous media.

\section{Green's functions of second order
elliptic differential operators} In application to the models
(2)-(7) we discuss the following equation for the Green's
functions
\begin{equation}
(B\partial_{0}^{2}+\partial_{j}A\partial_{j})G=2\delta
\end{equation}
where $0$ can be either a space coordinate, an imaginary time or
the real time. In the latter case $B$ should be negative.
 Let us  first discuss a special case $B=1$. Then,
\begin{equation}
(\partial_{0}^{2}+\partial_{j}A\partial_{j})G_{E}\equiv 2{\cal
A}G_{E}=2\delta
\end{equation}
 As an auxiliary tool for the wave equation we consider the
diffusion equation
\begin{equation}
\partial_{\tau}P=\frac{1}{2}\partial_{j}A\partial_{j}P
\end{equation}
We consider the fundamental solution $P_{\tau}$ of eq.(10) (the
transition function of a Markov process). The operator ${\cal A}$
defined on the lhs of eq.(9) is a non-negative symmetric operator
in the Hilbert space of square integrable functions. Hence, it has
the unique self-adjoint extension. We can define $\exp(-\tau{\cal
A})$ and subsequently $G_{E}$ as an integral kernel of
\begin{displaymath} {\cal
A}^{-1}=\int_{0}^{\infty}d\tau\exp(-\tau{\cal A})
\end{displaymath}
(the equality holds true on the domain where ${\cal A}^{-1}$ makes
sense;concerning Hilbert space methods for Green's functions see
\cite{maurin}). ${\cal A}$ is a sum of two commuting operators.
Hence, the kernel of $\exp(-\tau{\cal A})$ is a product of kernels
of these operators. In this way the solution of eq.(9) can be
expressed by $P_{\tau}$\begin{equation}
\begin{array}{l}
G_{E}(x_{0},{\bf x};x_{0}^{\prime},{\bf
x}^{\prime})=\int_{-\infty}^{\infty}d\omega\int_{0}^{\infty}d\tau
P(\tau,{\bf x},{\bf x}^{\prime})\exp(-\frac{1}{2}\omega^2\tau)
\exp(i\omega(x_{0}-x_{0}^{\prime})) \cr =\int_{0}^{\infty}d\tau
P(\tau,{\bf x},{\bf
x}^{\prime})(2\pi\tau)^{-\frac{1}{2}}\exp(-\frac{1}{2\tau}(x_{0}-x_{0}^{\prime})^{2})
\end{array}
\end{equation}
Let us denote by ${\bf q}_{s}({\bf x})$ the diffusion process
defined by the transition function $P$ (eq.(10)). Then, the
Green's function (8) can be expressed in the form
\begin{equation}
\begin{array}{l}
G_{E}(x_{0},{\bf x};x_{0}^{\prime},{\bf x}^{\prime})= \int
d\omega\int d\tau \cr
 E[ \exp\Big(-\omega^{2}\int_{0}^{\tau}B({\bf q}_{s}({\bf x}))ds\Big)
\exp(i\omega(x_{0}-x_{0}^{\prime}))\delta({\bf q}_{\tau}({\bf
x})-{\bf x}^{\prime})]
\end{array}
\end{equation}
where $E[..]$ denotes the expectation value over the diffusion
process $ {\bf q}_{\tau}$ (\cite{ikeda}\cite{simon}) defined by
the transition function $P$ of eq.(10) .
 We cannot find a general solution of eqs.(8)-(10)
 but restrict ourselves to
\begin{equation}
A( {\bf x})=k\vert {\bf x}\vert^{\alpha}
\end{equation}
where $k$ is a constant.

We can compute the transition function $P(\tau,{\bf 0},{\bf y})$
for a diffusion starting from a point ${\bf x}={\bf 0}$. Then, the
solution for $P$ in $D$ space-time dimensions is
\begin{equation}
P(\tau,{\bf 0},{\bf y})=K
(k\tau)^{-\gamma}\exp(-\frac{b}{2k\tau}\vert {\bf y}\vert^{\beta})
\end{equation}
where $K$ is a normalization constant resulting from $\int d{\bf
y}P=1$ and
\begin{equation}
\beta=2-\alpha
\end{equation}
\begin{displaymath}
b=4(2-\alpha)^{-2}
\end{displaymath}

 \begin{equation}
 \gamma=(D-1)(2-\alpha)^{-1}
 \end{equation}
The requirement of a normalization of $P$ imposes the condition
$\alpha<2$.

We can calculate the variance of the diffusion path
\begin{equation}
E[{\bf q}_{\tau}({\bf 0})^{2}]=\int d{\bf x}P(\tau,{\bf 0},{\bf
x}){\bf x}^{2}=C_{1}\tau^{\frac{2}{2-\alpha}}
\end{equation}
where $C_{1}$ is a constant (which can be calculated and depends
on the constants entering eq.(14), it is finite if $\alpha<2$).
Hence, for $\alpha>0$ we have a superdiffusive behaviour whereas
for $\alpha<0$ a subdiffusive one. There is no contradiction with
the diffusion equation (7) which determines the diffusion
coefficient as $lim_{\tau\rightarrow 0}E[\tau^{-1}q_{\tau}^{2}]$
because $A({\bf x})\rightarrow 0$ as ${\bf x}\rightarrow {\bf 0}$
in the superdiffusive case and $A({\bf x})\rightarrow \infty$ as
${\bf x}\rightarrow {\bf 0}$ in the subdiffusive behaviour. In
eq.(11) we can calculate directly the $\tau$ integral  if
$\gamma+\frac{1}{2}>1$. We obtain
\begin{equation}
G_{E}(x_{0},{\bf 0};x_{0}^{\prime},{\bf
x})=C_{2}\Omega_{E}(x_{0},{\bf 0};x_{0}^{\prime},{\bf
x})^{-\gamma+\frac{1}{2}}
\end{equation}with a certain constant
$C_{2}$ (depending on the parameters of eq.(14)) where
\begin{equation}
\gamma-\frac{1}{2}=(D-2+\frac{\alpha}{2})(2-\alpha)^{-1}
\end{equation}and
\begin{equation} \Omega_{E}(x_{0},{\bf 0};x_{0}^{\prime},{\bf
x})=(x_{0}-x_{0}^{\prime})^{2}+\frac{1}{k}(1-\frac{\alpha}{2})^{-2}\vert
{\bf x}\vert^{\beta}
\end{equation}
The function in the integral (11) is integrable at small $\tau $
for any $\gamma$ if $\Omega_{E}>0$ and for large $\tau$ if
$\gamma+\frac{1}{2}>1$, i.e. $D-2>-\frac{\alpha}{2}$. In the
limiting case of $\gamma=\frac{1}{2}$ the integral (11) is
logarithmically divergent for large $\tau$. The divergence means
that the operator ${\cal A}$ in eq.(9) is not positive definite.
We can still define the Green's function (9) in this limiting case
(as a distribution) subtracting the divergent constant from the
integral (11) (or imposing the condition $\int fd{\bf x}=0$ on the
test functions $f$). Then, the Green's function is proportional to
$\ln \Omega_{E}$.

 The
special case of eq.(9) is worth mentioning
\begin{equation}
\partial_{j}A\partial_{j}G_{E}=2\delta
\end{equation}
The solution is
\begin{equation}
\begin{array}{l}
G_{E}({\bf x},,{\bf x}^{\prime})=\int_{0}^{\infty}d\tau
P(\tau,{\bf x},{\bf x}^{\prime})
\end{array}
\end{equation}
The integral (22) is convergent if $\gamma>1$ ($D-1>2-\alpha$).
Inserting the explicit solution (14) we obtain
\begin{equation}
G_{E}({\bf 0},{\bf x})=C\vert {\bf x}\vert^{-D+3-\alpha}
\end{equation}
In the case $D-3=-\alpha$ again the logarithm will appear on the
rhs of eq.(23).

 The limiting case $\alpha\rightarrow 2$ is also
interesting because in this limit the formulas (14)-(20) lose
their meaning. In order to study what happens for $\alpha=2$ we
have found the diffusion process whose sample paths give the
transition function $P$ (solving eq.(10)). The square of it has
the form $({\bf q}_{\tau}({\bf x}))^{2}={\bf
x}^{2}\exp(c_{1}\tau+c_{2}b_{\tau})$ where $b_{\tau}$ is the
Brownian motion. It follows that ${\bf q}_{\tau}({\bf 0})={\bf 0}$
is the only solution starting from ${\bf 0}$ (there is the zero
solution also for $0<\alpha<2$ but in addition a non-trivial one
(14)). This result can explain why the formulae (14)-(18) have no
limit as $\alpha\rightarrow 2$. Now, instead of the power-law
behaviour (17) we obtain $E[ ({\bf q}_{\tau}({\bf x}))^{2}]={\bf
x}^{2}\exp(c\tau)$( where $c>0$ is a certain constant). We suggest
that when $\alpha\rightarrow 2$ then the power like relation
between $q$ and $\tau$ resulting from eqs.(14) and (17) is
replaced by an exponential one following from an exponential
increase of ${\bf q}_{\tau}^{2}$ (such a relation follows also
from an explicit calculation of the transition function for the
radial part of the process ${\bf q}_{\tau}$).

\section{The wave propagation and other applications}
We can continue eq.(20) into the real time . The result determines
the Feynman propagator \cite{dewitt}
\begin{equation}
G_{F}(t,{\bf 0};t^{\prime},{\bf
x})=C_{2}(\Omega_{F}+i\epsilon)^{-\gamma+\frac{1}{2}}
\end{equation}
where \begin{equation}
\Omega_{F}=-(t-t^{\prime})^{2}+\frac{1}{k}(1-\frac{\alpha}{2})^{-2}\vert
{\bf x}\vert^{\beta}
\end{equation}

 Let us note some characteristic features of this behaviour.
For ${\bf x}={0}$ we obtain
\begin{equation}
G_{F}(t,{\bf 0};t^{\prime},{\bf 0})= C_{3}\vert
t-t^{\prime}\vert^{-2(D-2+\frac{\alpha}{2})(2-\alpha)^{-1}}
\end{equation}
$G_{F}(t,{\bf 0};t^{\prime},{\bf 0})$ can have arbitrarily high
singularity (when $\alpha\rightarrow 2$). If $t=t^{\prime}$ then
\begin{equation}
G_{F}(t,{\bf x};t,{\bf 0})= C_{4}\vert{\bf
x}\vert^{-D+2-\frac{\alpha}{2}}
\end{equation}
Hence, the Green's function is integrable in ${\bf x}$. We can see
that for $\alpha>0$ the Green's function is more singular than the
 one with constant coefficients, whereas for $\alpha<0$ it is less singular.

The Feynman propagator does not appear in classical theory of wave
propagation (except of the Feynman-Wheeler electrodynamics)
because it propagates the waves from the past and from the future
as its real part is a sum of the advanced and retarded propagators
\begin{equation}
\Re G_{F}=\frac{1}{2}(G_{A}+G_{R})
\end{equation}
However, any solution $G$ of eq.(8) ($x_{0}=it$) is a sum
$G=G_{F}+U$ of $G_{F}$ and a solution $U$ of the wave equation
\begin{displaymath}
(B\partial_{t}^{2}-\partial_{j}A\partial_{j})U=0
\end{displaymath}
Choosing a regular U with some boundary conditions (in particular
at spatial infinity) we can obtain a Green's function with the
same singularity and appropriate boundary conditions.

 The real part of $G_{F}$ is a fractional derivative od
$\delta(\Omega_{F})$.
 If we
send a signal from ${\bf 0}$ at time $t$ then its arrival at time
$t^{\prime}$ at the point ${\bf x}$ can be calculated from the
equation $\Omega_{F}(t,{\bf 0};t^{\prime},{\bf x})=0$.  Let us
note that the wave front does not propagate with a constant speed
but the velocity depends on time (or space). So, we obtain from
$\Omega_{F}$
\begin{equation} \frac{d\vert{\bf x}\vert}{dt}=2\beta^{-1}
a^{\frac{1}{\beta}}(1-\frac{\alpha}{2})^{\frac{2}{\beta}}\vert
t-t^{\prime}\vert^{\frac{\alpha}{2-\alpha}}
\end{equation}
For the more general equation (8)
\begin{equation}
\begin{array}{l}
G_{E}(x_{0},{\bf x};x_{0}^{\prime},{\bf x}^{\prime})= \int d\tau
 E[ \exp\Big(-(x_{0}-x_{0}^{\prime})^{2}(4\int_{0}^{\tau}B({\bf q}_{s}({\bf
 x}))ds)^{-1}\Big)\cr
(4\pi\int_{0}^{\tau}B({\bf q}_{s}({\bf
 x}))ds)^{-\frac{1}{2}}\delta({\bf q}_{\tau}({\bf x})-{\bf x}^{\prime})]
\end{array}
\end{equation}
The case $\alpha=0$ in eq.(13) ($A=k$) is also interesting. Then,
${\bf q}_{s}({\bf x})={\bf x}+{\bf b}_{s}$, where ${\bf b}$
denotes the Brownian motion \cite{ikeda}\cite{simon}(in such a
case we know $P(\tau,{\bf x},{\bf x}^{\prime})$ explicitly).

The expectation value $E[.]$ in eq.(30) cannot be calculated
exactly. We have to resort to some approximations. If $B$ is a
bounded slowly varying function then approximately
\begin{equation}
B({\bf q}_{s}({\bf x}))\simeq B({\bf x})\simeq B({\bf
0})\end{equation} In such a case we obtain the formula (20) for
$G_{E}(x_{0},{\bf 0};x_{0}^{\prime},{\bf x}^{\prime} )$ with
\begin{equation} \Omega_{E}(x_{0},{\bf 0};x_{0}^{\prime},{\bf
x}^{\prime})=B({\bf
0})^{-1}(x_{0}-x_{0}^{\prime})^{2}+\frac{1}{k}(1-\frac{\alpha}{2})^{-2}\vert
{\bf x}^{\prime}\vert^{\beta}
\end{equation}
If $B$ like $A$ has a powerlike behaviour \begin{equation} B({\bf
x})=b\vert{\bf x}\vert^{\sigma} \end{equation} then the
approximation (31) cannot be justified. Let us note that from
eq.(14) it follows that
\begin{equation}
P(\tau,{\bf 0},{\bf y})=P(1,{\bf 0},\tau^{-\frac{1}{\beta}} {\bf
y})
\end{equation}
Hence,  \begin{equation} {\bf q}_{s}({\bf 0})\simeq
\tau^{\frac{1}{\beta}}{\bf q}_{\frac{s}{\tau}}({\bf 0})
\end{equation}
Inserting this approximate equality (this equality is exact for
the Brownian motion ${\bf b})$into eq.(30) and making the
approximation \begin{displaymath}\int_{0}^{1}B({ \bf q}_{s}({\bf
0}))ds\simeq E[\int_{0}^{1}B({ \bf q}_{s}({\bf
0})]ds=C\end{displaymath} we obtain
\begin{equation}
\begin{array}{l}
G_{E}(x_{0},{\bf 0};x_{0}^{\prime},{\bf
x})=KC^{-\frac{1}{2}}\int_{0}^{\infty}
d\tau\tau^{-\gamma-\frac{1}{2}-\frac{\sigma}{4-2\alpha}}\cr
\exp\Big(-\frac{1}{4C}\tau^{-1-\frac{\sigma}{2-\alpha}}
(x_{0}-x_{0}^{\prime})^{2}-\frac{b}{2k}\frac{\vert{\bf
x}\vert^{\beta}}{\tau}\Big)\end{array}
\end{equation}
We cannot compute the integral (36) exactly. Let us consider some
special cases. If $x_{0}=x_{0}^{\prime}$ then
\begin{equation}
G_{E}(0,{\bf 0};0,{\bf x})\simeq \vert{\bf
x}\vert^{-D+2-\frac{\alpha}{2}-\frac{\sigma}{2}}
\end{equation}
If ${\bf x}={\bf x}^{\prime}$
\begin{equation}
\begin{array}{l}
G_{E}(x_{0},{\bf 0};x_{0}^{\prime},{\bf 0})\simeq \vert
x_{0}-x_{0}^{\prime}\vert^{-\nu}\end{array}
\end{equation}
where
\begin{equation}
\nu=(2(D-2)+\alpha+\sigma)(2-\alpha+\sigma)^{-1}
\end{equation}
The case $\sigma=0$ and $\alpha>0$ could be considered as a
realistic approximation to eq.(2) describing the density of a
fluid concentrated at ${\bf x}=0$ ( then $\rho({\bf x})$ is
decreasing  from the origin as $\vert{\bf x}\vert^{-\alpha}$). In
 electrodynamics a scale invariant dielectric permittivity is
unrealistic. We could consider
\begin{equation}
\epsilon=\epsilon_{0}+b \vert {\bf x}\vert^{\alpha}
\end{equation}
where $\epsilon_{0}$ is a constant.
 In application to eq.(5)
   a source $\rho$ produces the electric potential
\begin{equation}
\phi({\bf y})=-\frac{1}{2}(G_{E}\rho)({\bf y})
\end{equation}
Hence, for a static point source located at ${\bf x}=0$ it follows
from eq.(23) that\begin{equation} \phi({\bf y})=C_{3}\vert {\bf
y}\vert^{3-D-\alpha}
\end{equation}
is valid for small distances if  $\alpha<0$ and for large
distances if $\alpha>0$ ($D=4$ in our world; eq.(42) modifies the
conventional $\frac{1}{r}$ Coulomb law). We obtain a similar
formula from eq.(4) for a magnetic field of a monochromatic wave.

The function $B$ in eq.(8) has a meaning of the magnetic
susceptibility for eq.(4). Hence, if
\begin{equation}
\mu=\mu_{0}+b\vert{\bf x}\vert^{\sigma}
\end{equation}
where $\mu_{0}$ is a constant then $\int_{0}^{\tau}B({\bf
q}_{s}({\bf 0}))ds\simeq \mu_{0}\tau
+c\tau^{1+\frac{\sigma}{\beta}}$. In such a case the
approximations made at eq.(30) hold true for small distances
(resulting from small $\tau$) if $\sigma$ is negative and large
distances if $\sigma $ is positive. The same argument holds true
when applied to eq.(3) with the dielectric permittivity of the
form (40). Then, the results of sec.2 concerning  scale invariant
coefficients apply to $\epsilon$ of the form (40)  if we consider
small distances in the case $\alpha<0$ and large distances if
$\alpha
>0$.

Finally, we  consider QED in an inhomogeneous medium. The quantum
electromagnetic field is defined by its time-ordered vacuum
correlation functions which are determined by the Feynman
propagator $G_{F}$. As an example of its applications we consider
 a scattering of photons on an external source $J$. The conventional
 formula for the $S$-matrix can be applied (see \cite{bialynicki})
 \begin{equation}
 S=\exp(\frac{i}{2}\int JG_{F}J)\exp(-i\int A^{(-)}J)\exp(-i\int
 A^{(+)}J)
 \end{equation}
 where $A^{(-)}$ is the part of the vector potential linear in the
 creation operators, whereas $A^{(+)}$ is the annihilation part
 (positive energy solution of the wave equation). In particular,
 if the source is of the form $J(t,{\bf x})=f(t)\delta({\bf x})$
 then the probability that  the source will cause no photon emission
 ( a preservation of the initial vacuum \cite{bialynicki})is
 \begin{equation}
 P(in)=\exp(-\int
 d\omega\vert\tilde{f}(\omega)\vert^{2}\tilde{G}_{F}(\omega))
 \end{equation}
 where tilde denotes the Fourier transform and $\tilde{G}_{F}(\omega)$ is the Fourier transform of the
 Green's function (38). If $f(t)=\cos(\omega_{0}t)$ then
 \begin{equation}
 \ln P(in)\simeq \omega_{0}^{\nu-1}\delta(0)\simeq
 T\omega_{0}^{\nu-1}
 \end{equation}
 where $\delta(0)$ is interpreted as the duration $T$of the signal
 $J$ (this interpretation follows from the integral
 $\int JG_{F}J$ in eq.(44) if $T$ is large in comparison to the period
 $\frac{2\pi}{\omega_{0}}$ ). We can see that the inhomogeneity
 of the medium could be measured by a photon counting experiments
 showing whether $\omega ^{\nu-1}$ is linear in $\omega$ or not.
\section{Summary}
The potential of a point charge is determined by the Poisson
equation (5) with $\rho $ as a $\delta$ source. The electric field
can be obtained from eq.(3). The spatial singularity and the
decrease at infinity of the potential and of the electric field
are  determined by the behaviour of the Green's functions. We have
shown that this behaviour is not universal and may depend on the
properties of the medium in which the charge is embedded.
Analogous formulae are known for acoustics and fluid dynamics
expressing the intensity of the waves produced by a given source.
The laws governing the production of such waves will be modified
if the fluid density or its compressibility is a growing or
singular function of ${\bf x}$.
 In our examples  there is
always a distinguished point in $G_{E}({\bf x},{\bf x}^{\prime}) $
identified as an origin
 ${\bf x}=0$
of the coordinate system. This point can be distinguished by the
location of the source. The special role of the origin will
disappear if coefficients in the differential operators are random
variables invariant under translations ( we discussed such random
Green's functions in \cite{habaran}).

{\bf Acknowledgements}

The author thanks the anonymous referees for valuable remarks
 
\end{document}